\documentclass[12pt,preprint]{aastex}

\newcommand{\be}{\begin{equation}}
\newcommand{\ee}{\end{equation}}
\def\lta{\,\raise 0.3 ex\hbox{$ < $}\kern -0.75 em
 \lower 0.7 ex\hbox{$\sim$}\,}
\def\gta{\,\raise 0.3 ex\hbox{$ > $}\kern -0.75 em
 \lower 0.7 ex\hbox{$\sim$}\,} 
\newcommand{\mnine}{m_9}
\newcommand{\anine}{a_9}
\newcommand{\enine}{e_9}
\newcommand{\sigmaint}{\sigma_{\rm int}}

\begin{document} 

\title{Interaction Cross Sections and Survival Rates\\
for Proposed Solar System Member Planet Nine} 

\author{Gongjie Li$^1$ and Fred C. Adams$^{2,3}$}

\affil{$^1$Harvard-Smithsonian Center for Astrophysics, 
The Institute for Theory and Computation, \\
60 Garden Street, Cambridge, MA 02138, USA} 

\affil{$^2$Physics Department, University of Michigan, Ann Arbor, MI 48109, USA} 

\affil{$^3$Astronomy Department, University of Michigan, Ann Arbor, MI 48109, USA} 

\begin{abstract}  

Motivated by the report of a possible new planetary member of the
Solar System, this work calculates cross sections for interactions
between passing stars and this proposed Planet Nine. Evidence for the
new planet is provided by the orbital alignment of Kuiper Belt
objects, and other Solar System properties, which suggest a
Neptune-mass object on an eccentric orbit with semimajor axis
$\anine\approx400-1500$ AU. With such a wide orbit, Planet Nine has a
large interaction cross section, and is susceptible to disruption by
passing stars.  Using a large ensemble of numerical simulations
(several million), and Monte Carlo sampling, we calculate the cross
sections for different classes of orbit-altering events: [A]
scattering the planet into its proposed orbit from a smaller orbit,
[B] ejecting it from the Solar System from its current orbit, [C]
capturing the planet from another system, and [D] capturing a
free-floating planet.  Results are presented for a range of orbital
elements with planetary mass $\mnine=10M_\earth$. Removing Planet Nine
from the Solar System is the most likely outcome. Specifically, we
obtain ejection cross sections $\sigmaint\sim5\times10^6$ AU$^2$
($5\times10^4$ AU$^2$) for environments corresponding to the birth
cluster (field). With these cross sections, Planet Nine is likely to
be ejected if the Sun resides within its birth cluster longer than
$\Delta{t}\gta100$ Myr. The probability of ejecting Planet Nine due to
passing field stars is $\lesssim 3\%$ over the age of the Sun.
Probabilities for producing the inferred Planet Nine orbit are low
$(\lta5\%)$.

\end{abstract} 


\bigskip
$\,$
\bigskip
$\,$ 
\bigskip

\section{Introduction} 
\label{sec:intro} 

Evidence for the existence of a new large planet in our Solar System
has recently been reported \citep{phattie}. This body, given the
working name Planet Nine, provides a viable explanation for the
orbital alignment observed for a collection of Kuiper Belt objects on
wide orbits, the existence of detached Sednoids, and highly inclined
Centaurs with large semimajor axes \citep{ts2014,dlfm}.  To explain
these Solar System properties, the hypothetical Planet Nine must have
mass $\mnine\approx10M_\earth$, semimajor axis $\anine\approx400-1500$
AU, and eccentricity $\enine\approx0.4-0.9$.  Given the potential
importance of a massive new Solar System member, any constraints on
its orbital properties, dynamical history, and/or formation mechanism
are of great interest.  With its wide orbit, Planet Nine is highly
susceptible to gravitational perturbations from passing stars and
binaries \citep{al2001,adams2006,malm2007,malm2011,veras}. The goal of
this paper is to calculate the cross sections for interactions between
the Solar System and passing stars, both in the environment of the
Solar birth cluster and in the field. These cross sections can then be
used to estimate probabilities for Planet Nine to survive, or to be
captured into its elongated orbit.

Astronomy has a long history of considering the gravitational
perturbations due to previously undetected Solar System bodies acting
on known objects. This course of action led to the prediction and
successful detection of the planet Neptune \citep{lever,galle,jcadams}.  
Later on, the existence of a large Planet X was predicted to reside
outside Neptune. Although the original Planet X was never found, the
subsequent search ultimately led to the detection of the dwarf planet
Pluto (see \citealt{tombaugh} for more historical detail). However,
not all such predictions have been fruitful: Nemesis, a hypothetical
red dwarf companion to the Sun with semimajor axis $a_N\sim10^5$ AU
\citep{nemesis}, was invoked to explain periodic extinctions in the
fossil record, and provides a cautionary example.\footnote[2]{The 
results of this paper indicate that Nemesis would have a negligible
chance of survival in its proposed distant orbit.} As outlined below,
this paper places significant constraints on possible formation
scenarios and dynamical histories for Planet Nine.

Dynamical scattering interactions between the Solar System and passing
stars can take place over a wide range of parameter space. This paper
focuses on interactions that affect the orbit of Planet Nine, which we
consider to be dynamically decoupled from the rest of the Solar System.  
Specifically, the giant planets (Jupiter through Neptune) are too far
inside the orbit of Planet Nine to affect its dynamics during close
encounters, whereas Kuiper Belt objects have too little mass.  The
results of this paper support this decoupling hypothesis (Section
\ref{sec:csection}).

Here we consider two background environments for scattering
interactions: the Solar birth cluster and the field. Most stars form
within stellar clusters of some type \citep{ladalada,porras}. Based on
its observed properties, including radioactive enrichment levels and
well-ordered orbits, our Solar System is likely to have formed within
a cluster containing $N_\ast=10^3-10^4$ members
\citep{al2001,zwart,adams2010,pfalzner}; these same considerations
suggest that the time spent in the birth cluster is $\sim100$ Myr (see
also \citealt{pfalznerrev}).  Within the cluster, with relative
velocity dispersion $v_b\sim1$ km/s, the Sun samples a range of
stellar densities, with mean $\langle{n_\ast}\rangle\sim100$
pc$^{-3}$.  After leaving its birth environment, the Solar System has
lived an additional $\sim4.5$ Gyr in the Solar neighborhood, which is
characterized by stellar density $\langle{n_\ast}\rangle\sim0.1$
pc$^{-3}$ and velocity $v_b\sim40$ km/s \citep{bintrem}. Although the
Solar System spends more time in the diffuse environment of the field,
interactions are more likely in the Solar birth cluster due to the
increased stellar density and larger cross sections (smaller $v_b$).

Several types of scattering interactions are of interest. First,
because planet formation is difficult on wide orbits \citep{dodson},
Planet Nine could be formed in a nearly circular orbit with a much
smaller semimajor axis ($a_0\ll{a_9}\sim400-1500$ AU), and/or
scattered by the giant planets into an orbit with perihelion
$p\sim30-40$ AU \citep{al2003,veras}.  The window for scattering a
planet into an orbit with $a\sim1000$ AU is narrow (the required
energy change is a large fraction of the initial binding energy, so
that most planets are ejected instead of retained at large semimajor
axis). In any case, we want to calculate cross sections for scattering
Planet Nine from such initial orbits into its proposed long-term
state. Second, after its formation, while Planet Nine traces through
its distant orbit, it can be removed from the Solar System by passing
stars. We thus need to calculate the cross sections for planet
ejection. Finally, Planet Nine, and its orbit, could be produced
through a capture event. We thus need the cross sections for capturing
Planet Nine from another solar system, and/or as a freely-floating
planet within the Solar birth cluster.

This paper is organized as follows. Section \ref{sec:csection} reviews
the technique used to calculate interaction cross sections and
presents results for the channels of scattering interactions outlined
above. These cross sections are used in Section \ref{sec:probability} to
estimate probabilities for producing Planet Nine by scattering from an
internal orbit, capturing it from another solar system, capturing it
from a freely-floating state, and ejecting Planet Nine from the Solar
System. The paper concludes in Section \ref{sec:conclude} with a
summary of our results and a discussion of their implications.

\section{Interaction Cross Sections} 
\label{sec:csection} 

In order to calculate the cross sections for interactions between the
Solar System and passing stars, we perform a large ensemble of
numerical experiments using previously developed techniques (for
details, see \citealt{al2001,gdawg}).  This approach is summarized
below:

The first step is to specify the initial configuration of the Solar
System. In this context, we are interested in the related issues of
producing the desired orbit of Planet Nine via scattering
interactions, removing Planet Nine from its wide orbit, and capturing
Planet Nine into the Solar System.  To study the first issue, the
initial configurations are circular orbits with semimajor axes
$a_0=200,400,600$ and 800 AU. Note that formation of Planet Nine would
be difficult for the larger semimajor axes, but interaction cross
sections are small for the lower values.  For ejection events, we
consider starting states with $a_9=400,600,800,$ and 1000 AU, and
orbital eccentricities $e_0=0.4,0.6,$ and 0.8. For capture events, we
consider the Sun to interact with a passing star that harbors Planet
Nine. The initial orbit is assumed to be circular with semimajor axes
$a_0=50-200$ AU.  We consider initial stellar hosts with
$M_\ast=1M_\odot$, $M_\ast=0.3M_\odot$, and with a range of masses
sampling the stellar IMF. We also consider capture events where Planet
Nine starts as a freely-floating body in the Solar birth cluster
\citep{perets}. The mass of Planet Nine is taken to be
$\mnine=10M_\earth\ll{M_\ast}$ for all simulations.

Next we specify the background environment, i.e., the Solar birth
cluster or the field. The background determines the velocity
distribution from which the encounter speeds are sampled.  This
distribution is assumed to be Maxwellian with dispersion $v_b=1$ km/s
for the cluster environment \citep{ladalada,porras} and $v_b=40$ km/s
in the solar neighborhood \citep{bintrem}. Here we use the  
distribution of encounter velocities, which has a wider Maxwellian
form with relative dispersion $\sqrt{2}v_b$. In addition, the velocity 
distribution used in our expectation values includes an additional
factor of $v$ so that we calculate cross sections
$\sigmaint\equiv\langle{v\sigma}\rangle/v_b$, rather than
$\langle{v\sigma}\rangle$ \citep{bintrem}.  Note that the stellar
density determines the interaction rates and the corresponding
probabilities (Section \ref{sec:probability}), but does not affect
calculations of the cross sections.

For the interacting stars, we consider binary systems because a
sizable fraction of stars are found in binaries and because they
produce significantly larger scattering effects. The binary periods,
eccentricities, and mass ratios are sampled from observed
distributions \citep{duque}; the primary mass is sampled from a
standard log-normal form for the stellar initial mass function. The
phases of the orbits are uniform-random and the angles that determine
the geometry of the interaction are chosen to produce an isotropic
distribution of appoach directions. Finally, we must specify the
impact parameter $h$ of the interactions. In principle, one should
consider the full range of impact parameters out to $h\to\infty$; in
practice, however, distant encounters have little effect and do not
contribute to the cross sections.  Here we adopt a maximum allowed
impact parameter $h_{max}=30a_b$, where $a_b$ is the semimajor axis of
the binary. This varying distribution of impact parameters must be
taken into account in determining the cross sections \citep{al2001}.
Because the impact parameter has a maximum value, the resulting cross
sections represent lower limits to their true values.

With the starting states and interaction environment specified, we
perform a large ensemble of numerical simulations, where the relevant
parameters are sampled from the distributions described above using a
Monte Carlo scheme. For each starting configuration of the Solar
System and each choice of background environment, we perform
$N\approx500,000$ numerical simulations, where the initial conditions
for each case represent an independent realization of the parameters.
The results are then used to calculate the interaction cross sections
for the events of interest.

\bigskip

\begin{table}[tbh]\label{table1}
\centerline{\bf Cross Sections for Formation, Ejection, and Capture of Planet Nine} 
\centerline{$\,$} 
\begin{center}
\begin{tabular}{lccccc}
\hline 
\hline
\hline 
Event Class & $e_0$ & $\sigmaint(a_1)$ & $\sigmaint(a_2)$ & $\sigmaint(a_3)$ & $\sigmaint(a_4)$ \\
$\,$ & $\,$ & (AU$^2$) & (AU$^2$) & (AU$^2$) & (AU$^2$) \\
\hline 
\hline               
Formation & 0 &  94600 $\pm$ 3400 & 485200 $\pm$ 8300 & 1020700 $\pm$ 12900  & 1261000  $\pm$ 14500 \\ 
\hline                         
Ejection & 0.4 & 1907900 $\pm$ 16200 & 2656700 $\pm$ 19400 & 3090800 $\pm$ 21500 & 3879400 $\pm$ 24400 \\
Ejection & 0.6 & 2645700 $\pm$ 19600 & 3811000 $\pm$ 24400 & 4433400 $\pm$ 26600 & 5782600 $\pm$ 30500 \\                         
Ejection & 0.8 & 4834000 $\pm$ 28600 & 7017900 $\pm$ 35800 & 9386900 $\pm$ 40700 & 12226800 $\pm$ 48400 \\                
\hline                                                                    
Ejection-field & 0.4 & 12000 $\pm$  1100 & 15900 $\pm$  1200  & 27500  $\pm$  1500 & 33800 $\pm$ 1800\\                         
Ejection-field & 0.6 & 15600 $\pm$  1200 & 20800 $\pm$  1400 & 41900 $\pm$ 2000 & 20800 $\pm$ 2100 \\                           
Ejection-field & 0.8 & 25600 $\pm$  1500 & 32100 $\pm$  1600 &  62500 $\pm$  2500 & 64600 $\pm$ 2600\\                           
\hline 
Capture(IMF) & 0 & 29400 $\pm$ 1400 & 55100 $\pm$ 2000 & 72000 $\pm$ 2400 & 85600 $\pm$ 2600\\                                   
Capture(m=1) & 0 & 26800 $\pm$ 1300 & 47500 $\pm$ 2000 &  74900 $\pm$ 2500 & 88300  $\pm$ 2600\\                                     
Capture(m=0.3) &0 & 29200 $\pm$ 1400 & 54900 $\pm$ 2000 & 78100 $\pm$ 2400 &  97800 $\pm$ 2900\\                                    
Capture(IMF) & 0.6 & 32700 $\pm$ 1600 & 57700 $\pm$ 2500 & 72900 $\pm$ 2600 & 85000 $\pm$ 3200\\                                
\hline
Capture(f$\to$b) & $-$ & 11800 $\pm$ 1300 & 16900 $\pm$ 1500 & 22000 $\pm$ 1700 & 79000 $\pm$ 2800 \\ 
\hline
\hline
\hline
\end{tabular}
\vspace{0.25cm}
\caption{For each type of interaction (labeled in left column) cross 
sections are given for four values of semimajor axis (which vary with
the type of event).  Errors represent uncertainties due to incomplete
Monte Carlo sampling. For formation events, Planet Nine starts in our
Solar System with an orbit at $a_0=200,400,600,800$ AU.  For ejection
events, Planet Nine starts in a circumsolar orbit with
$a_9=400,600,800,1000$ AU. For capture events from bound orbits,
Planet Nine starts in another solar system with $a_0=50,100,150,200$
AU. For the capture of freely-floating planets (f$\to$b), cross
sections are given for final (circumsolar) orbits with
$a_f<a_f$(max)$=1000,1500,2000$ AU, and $\infty$. }
\end{center}
\end{table}  

The interaction cross sections are listed in Table~1 for the different
classes of scattering events. These cross sections listed are 
the expectation values $\sigmaint=\langle{v\sigma}\rangle/v_b$, where
the angular brackets denote averaging over the distribution of
encounter speeds. First we consider cross sections for the formation
of the Planet Nine orbit from an interior launching location.  For the
cross sections listed, the planet starts in a circular orbit with a
range of starting semimajor axes, and the Solar System is assumed to
reside in its birth cluster ($v_b=1$ km/s).  The requirements for
successful production of the Planet Nine orbit are that the final
semimajor axis lies in the range $400<a_f<1500$, with orbital
eccentricty $0.4<e_f<0.9$ and inclination angle $i_f<60^\circ$. The
resulting cross sections are listed in the first line of Table~1 for
four starting values of semimajor axis $a_0=200,400,600,$ and 800
AU. These cross sections for producing the required Planet Nine orbit
in a cluster environment are approximately the geometric cross
section of the original orbit (specifically, this ratio lies in the
range $0.6-1$).  For completeness, we have calculated cross
sections for producing Planet Nine from smaller initial
orbits. However, the resulting cross sections are small, 
$\sigmaint=133,974,3181,$ and $13053$ AU$^2$ for $a_0=10,30,50,$ and 100
AU.  Finally, given that Planet Nine could have formed in the giant
planet region and been scattered outward, we have calculated cross
sections for producing Planet Nine from eccentric orbits with
$a_0=50-200$ AU and perihelion $p=a(1-e)=5-40$ AU. However, these
cross sections are somewhat smaller than those listed in Table 1 for
circular starting orbits.

Next we consider the cross sections for the ejection of Planet Nine.
For this class of events, we consider six ensembles of simulations,
with three starting values of eccentricity ($e_9=0.4,0.6,0.8$) and two
velocity scales ($v_b$ = 1 and 40 km/s, for the Solar birth cluster
and the field). Here we consider four possible values for the
semimajor axis of Planet Nine $a_9=400,600,800,$ and 1000 AU (see
Table~1). For interactions in a low-speed environment ($v_b=1$ km/s),
the cross sections for ejection of Planet Nine are approximately an
order of magnitude larger than the cross sections for scattering
Planet Nine into its orbit (see Table~1).\footnote[2]{Note that the  
starting semimajor axes $a_0$ for formation events are not the same as
the initial values $a_9$ for ejection events.}  This finding suggests
that the production of Planet Nine is unlikely to take place by
scattering from a close orbit into its required wide orbit.

\begin{figure} 
\figurenum{1} 
{{\epsscale{0.80} \plotone{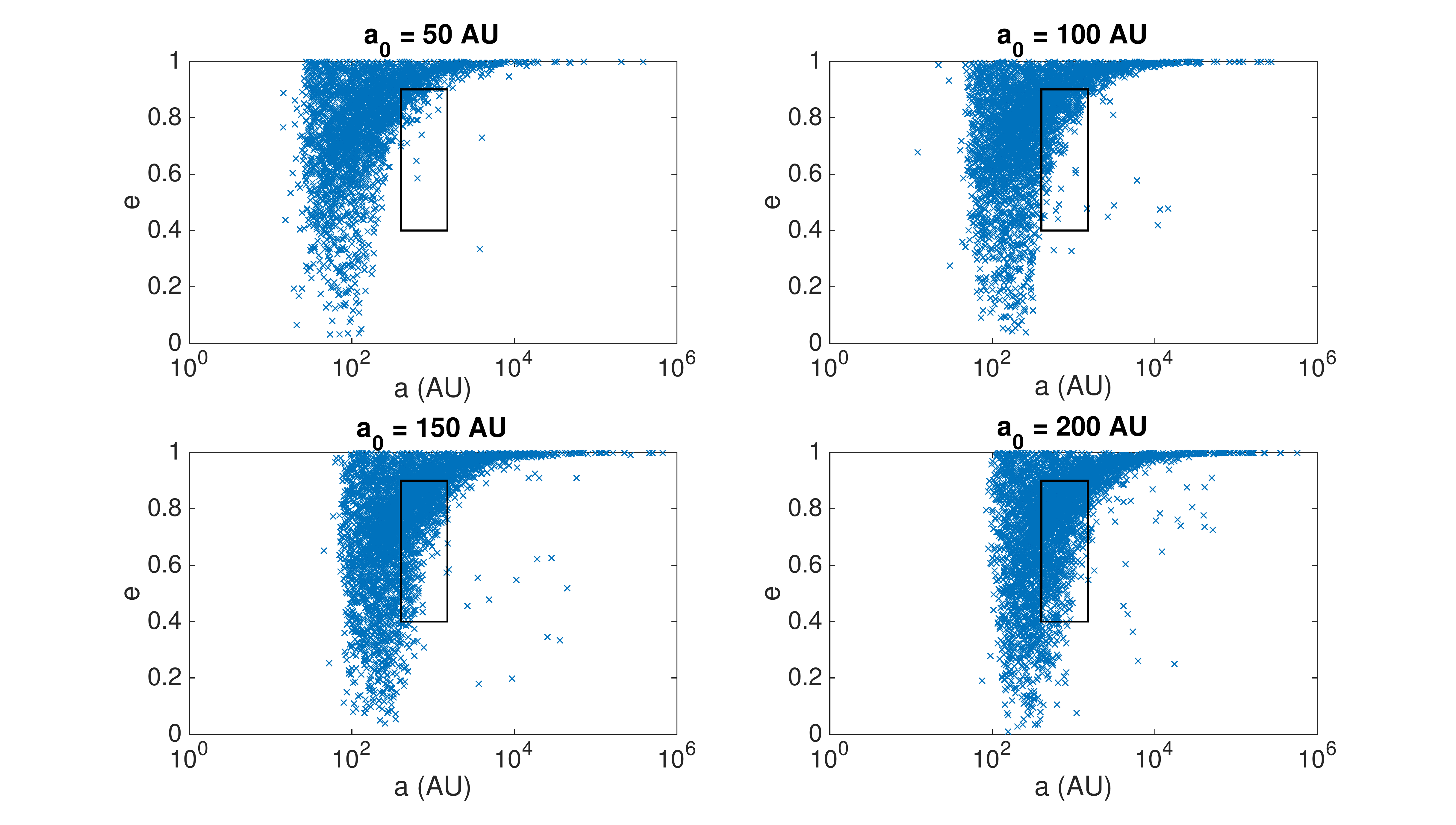} } } 
\figcaption{Post-encounter orbital elements for Planet Nine analogs  
that are captured by the Sun from other solar systems.  The panels
show results for four values of the starting semimajor axes ($a_0$ =
50, 100, 150, and 200 AU) for the planet, which initially belongs to
another star in a circular orbit ($e_0=0$). Black boxes delineate 
acceptable orbital elements. } 
\label{fig:aecapture} 
\end{figure}  

In principle, Planet Nine could be captured from another solar system.
To address this issue, we first assume that the planet initially
resides in orbit around another star. For these interactions, the
passing ``binary'' is thus the host star with Planet Nine as its
companion.  The starting orbit of Planet Nine is taken to have
semimajor axes $a_0=50,100,150$, and 200 AU and vanishing eccentricity
$e_0=0$. The interactions take place in the Solar Birth cluster with
$v_b=1$ km/s.  We consider the capture of Planet Nine from solar
systems with several types of host stars. First, the mass of the star
is sampled from the stellar IMF; we also consider cases where the host
star has fixed mass $M_\ast=1M_\odot$ and $M_\ast=0.3M_\odot$. For
comparison, we also consider initial orbits with non-zero eccentricity
$e_0=0.6$.

The resulting cross sections for the Sun to capture Planet Nine from
another star are given in Table~1. These capture cross sections are
much smaller than the ejection cross sections within the Solar birth
cluster, but are roughly comparable to ejection cross sections in the
field (where $v_b=40$ km/s).  The capture cross sections for M-dwarf
host stars ($M_\ast=0.3M_\odot$) are nearly the same as those
calculated by sampling the stellar IMF, whereas cross sections for
$M_\ast=1M_\odot$ hosts are mostly smaller. Finally, the cross
sections for capture from initial orbits with $e_0=0.6$ are comparable to
  those with $e_0=0$.

These capture cross sections (listed in Table~1) include all events
where the Sun successfully steals Planet Nine from another star, into
any bound orbit.  However, not all capture events lead to acceptable
orbits for newly acquired planet. Figure \ref{fig:aecapture} shows the
post-encounter orbital elements for Planet Nine, in its new orbit
about the Sun, for four values of the starting semimajor axis (where
the host-star mass is sampled from the IMF). Capture events produce
final-state orbits with the full distribution of eccentricities,
$0\le{e}\le1$, so that the target range for Planet Nine
$(0.4\le{e}\le0.9)$ is readily obtained (see also \citealt{jilkova}).
The distribution of final-state semimajor axes is also wide, where the
target range $a_f=400-1500$ AU is often realized. Specifically, the
fractions of acceptable final state orbits (out of the total number of
capture events) are 3.2\% ($a_0=50$ AU), 14\% ($a_0=100$ AU),
24\% ($a_0=150$ AU), and 32\% ($a_0=200$ AU).

Finally, we consider the capture of free-floating planets. For these
simulations, the Solar System starts with its four giant planets in
their current orbits, and then interacts with a free-floating Planet
Nine. The total cross sections for capturing freely-floating planets
(final line of Table 1) are similar to those for capturing bound
planets. However, the fraction of acceptable final states is only
$\sim8.4\%$, so it is more likely to capture a bound planet
with large $\gtrsim 100$ AU orbit.

With the cross sections determined, we can assess our assumption that
the Planet Nine orbit is dynamically decoupled from the inner Solar
System. The giant planet orbits are vulnerable to changes during
scattering encounters, and the cross sections for their disruption
have been calculated previously \citep{al2001,gdawg}.  Specifically,
the cross section for doubling the eccentricity of Neptune, or the
spread in inclination angles, is $\sigma_{dis}\approx$ (400 AU)$^2$.
This cross section is thus smaller than that required to eject Planet
Nine by $1-2$ orders of magnitude, depending on $\anine$ (Table 1). As
a result, severe changes to the Planet Nine orbit (ejection) are much
more likley than modest changes to the giant planet orbits.

Although the giant planets are relatively unperturbed, the wider
orbits of KBOs can be altered during scattering encounters. We have
performed additional ensembles of simulations to address this issue:
First, we placed KBOs in orbits at $a_0=200-500$ AU in simulations
where the Solar System captures Planet Nine as a freely-floating
planet. For KBO orbits with initial eccentricity $e_0=0.7$, the
post-encounter eccentricities always fall within the range
$e_f=0.55-0.8$. As a result, Planet Nine capture events produce only
moderate eccentricity changes in the wide-orbit KBO population. In
contrast, if Planet Nine is captured from a bound orbit in another
Solar System, the KBOs can be scattered significantly. Specifically,
the cross sections for ejecting KBOs ($a_0=200-500$ AU, $e_0=0.7$) are
somewhat larger than the cross sections for capturing Planet Nine. 
Finally, we note that KBO orbits can also be re-populated through
close stellar encounters \citep{kenyon}. 

The simulations of this paper are limited to 40 dynamical  
times of the initial (separated) systems ($t\lta10^5$ yr). This
interval includes many dynamical times of the giant planets, but 
does not capture long-term secular interactions (see also 
\citealt{gdawg}). The cross sections presented here thus describe the
immediate changes in the orbital elements, whereas additional changes
can occur over longer time scales. Since Planet Nine is thought to
reside in an inclined orbit, e.g., it could enforce Kozai-Lidov
oscillations of the giant planets over secular time scales.  These
longer term effects, which should be considered in future work, are
potentially significant for any scenario that includes Planet Nine.

\section{Scattering and Survival Probabilities} 
\label{sec:probability} 

Given the interaction cross sections calculated in Section
\ref{sec:csection} for different scattering events, we now estimate
the probabilities of those events occurring. The scattering rate is
given by 
\be\Gamma=n_\ast\langle{v\sigma}\rangle=n_\ast\sigmaint{v_b},\label{intrate}\ee 
where $\sigmaint$ is the cross section for the particular event, 
and the corresponding optical depth for scattering has the form 
\be{P}=\int{dt}n_\ast\sigmaint{v_b}.\label{opdepth}\ee 
For $P<1$, this quantity defines the probability of scattering; for
larger values $P>1$, the probability $\cal{P}$ for survival in the
face of scattering has the form $\cal{P}$=$\exp[-P]$.  Because the
velocity dependence of scattering cross sections has the approximate
power-law form $\sigmaint\propto{v^{-7/5}}$ \citep{gdawg}, the product
$v\sigmaint\propto{v^{-2/5}}$, so that the integrand in equation
(\ref{opdepth}) does not depend sensitively on the time-dependence of
the velocity. We then write equation (\ref{opdepth}) in the form 
\be{P}=\sigmaint(v_b)\int{dt}n_\ast{v_b}\equiv\sigmaint(v_b)(\Delta{t})v_b\langle{n_\ast}\rangle,\ee
where $v_b$ is the velocity dispersion of the environment and the
final equality defines the time-and-velocity-averaged density
$\langle{n_\ast}\rangle$ of the background. 

First we consider the properties of the Solar birth aggregate.
Previous considerations have placed constraints on this environment
and indicate that the Sun was born within a cluster of intermediate
stellar membership size $N_\ast\sim3000$
\citep{adams2010,pfalzner,zwart} where the velocity scale $v_b=1$
km/s. These same considerations suggest that the Solar System is
likely to reside in the cluster for an integration time
$(\Delta{t})\approx100$ Myr, a typical value for the lifetime of an
open cluster. The mean stellar density in clusters in the solar
neighborhood \citep{ladalada} is nearly constant as a function of
stellar membership size $N_\ast$ (see Figure 1 of \citealt{proszkow}),
and has a value $\langle{n_\ast}\rangle\approx100$ pc$^{-3}$, which is
used to calculate probabilities for this paper.

We also consider the Solar neighborhood in the Galaxy, where the
velocity dispersion $v_b=40$ km/s \citep{bintrem} and the integration
time $(\Delta{t})=4.6$ Gyr.  The stellar density at the solar circle
is difficult to measure. Here we use recent estimates where
$\langle{n_\ast}\rangle\approx0.1$ pc$^{-3}$ \citep{mckee,bovy}. Note
that the velocity dispersion of stars tends to increase with stellar
age \citep{nordstrom}, although this complication is not included.  
As outlined above, the scattering rate
$\Gamma\propto(v\sigmaint)\propto{v}^{-2/5}$ varies slowly with velocity.
We expect $v_b$ to vary by a factor $\lta2$ over the age of the Solar
System in the field, so that the probability estimate changes by a factor $\lta1.3$. 
\begin{figure} 
\figurenum{2} {{\epsscale{0.80} \plotone{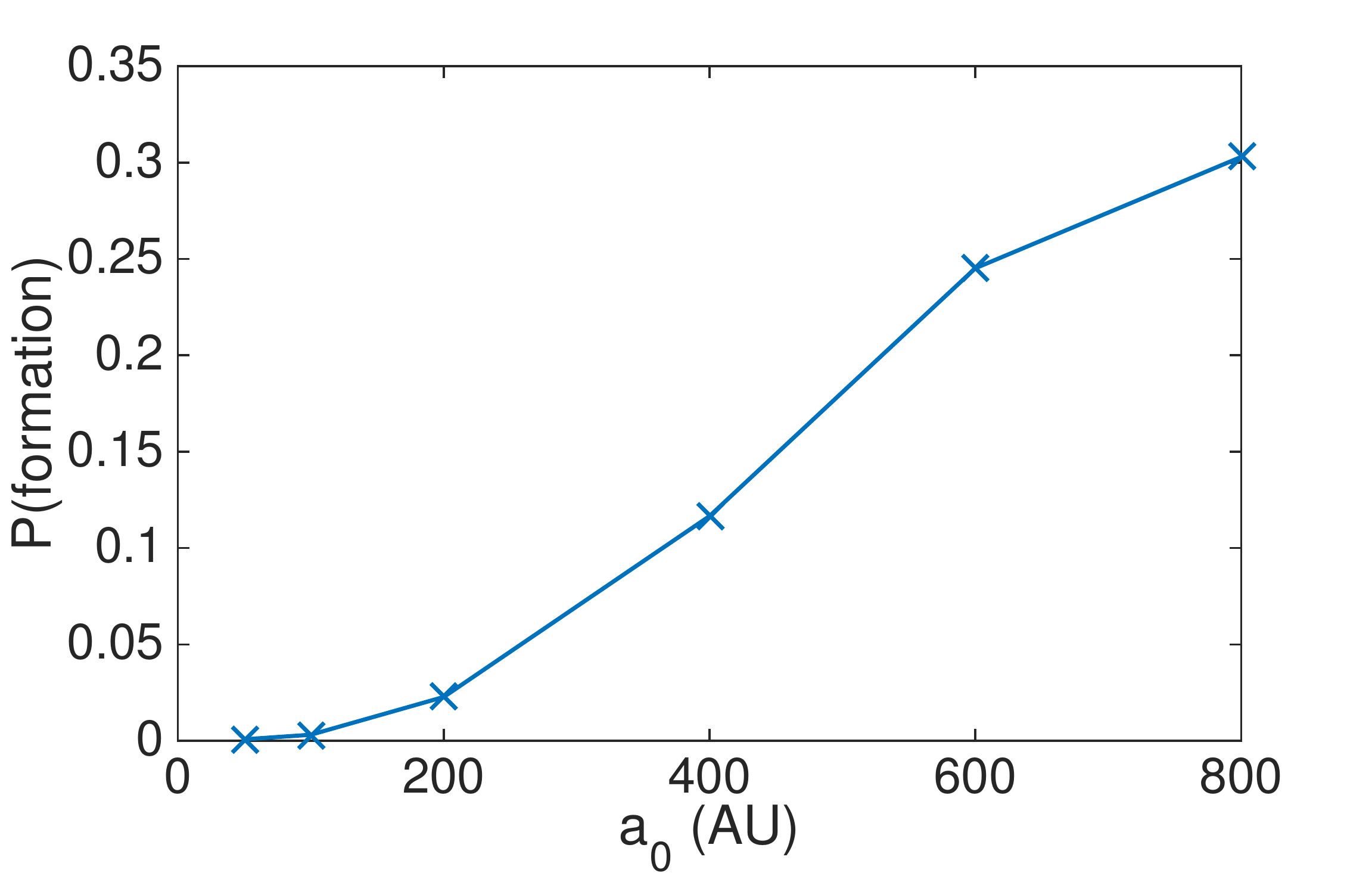} } }
\figcaption{Scattering optical depths for producing the required 
Planet Nine orbit, where the planet starts with smaller semimajor axis
$a_0$ and $e_0=0$. The final-state orbital elements are required to
fall in the allowed ranges 400 AU $\le{a_f}\le$ 1500 AU, 0.4
$\le{e_f}\le$ 0.9, and $i_f\le60^\circ$. } 
\label{fig:formation} 
\end{figure} 

The first class of interactions under consideration is the possible
production of Planet Nine orbital elements from less extreme initial
conditions. Here we take the starting configurations to be circular
orbits with a range of semimajor axes $a_0=200-800$ AU. The final
system parameters that represent successful production require orbital
eccentricity $0.4<e_f<0.9$, semimajor axis $a_f=400-1500$ AU, and
inclination angle $i_f<60^\circ$ \citep{phattie}.  Figure
\ref{fig:formation} shows the scattering optical depths for producing
Planet Nine as a function of the starting semimajor axis.  The
formation probabilties are low unless the starting semimajor axis is
large, $a_0\gta400$ AU, comparable to the required final values of
$a_9$. Even for $a_0=600$ AU, the preferred final value, the
probability of scattering Planet Nine into its required orbit is only
$\sim25\%$. The probabilities fall to $\sim3\%$ for $a_0=200$ AU.
Planets starting with orbits in the giant planet region of our Solar
System have a negligible chance of scattering into the required
orbit. This finding is consistent with previous results
\citep{al2001,gdawg}, which show that eccentricity and inclination
angles are much more readily altered than semimajor axes during 
scattering interactions.

\begin{figure} 
\figurenum{3} 
{{\epsscale{0.90} \plotone{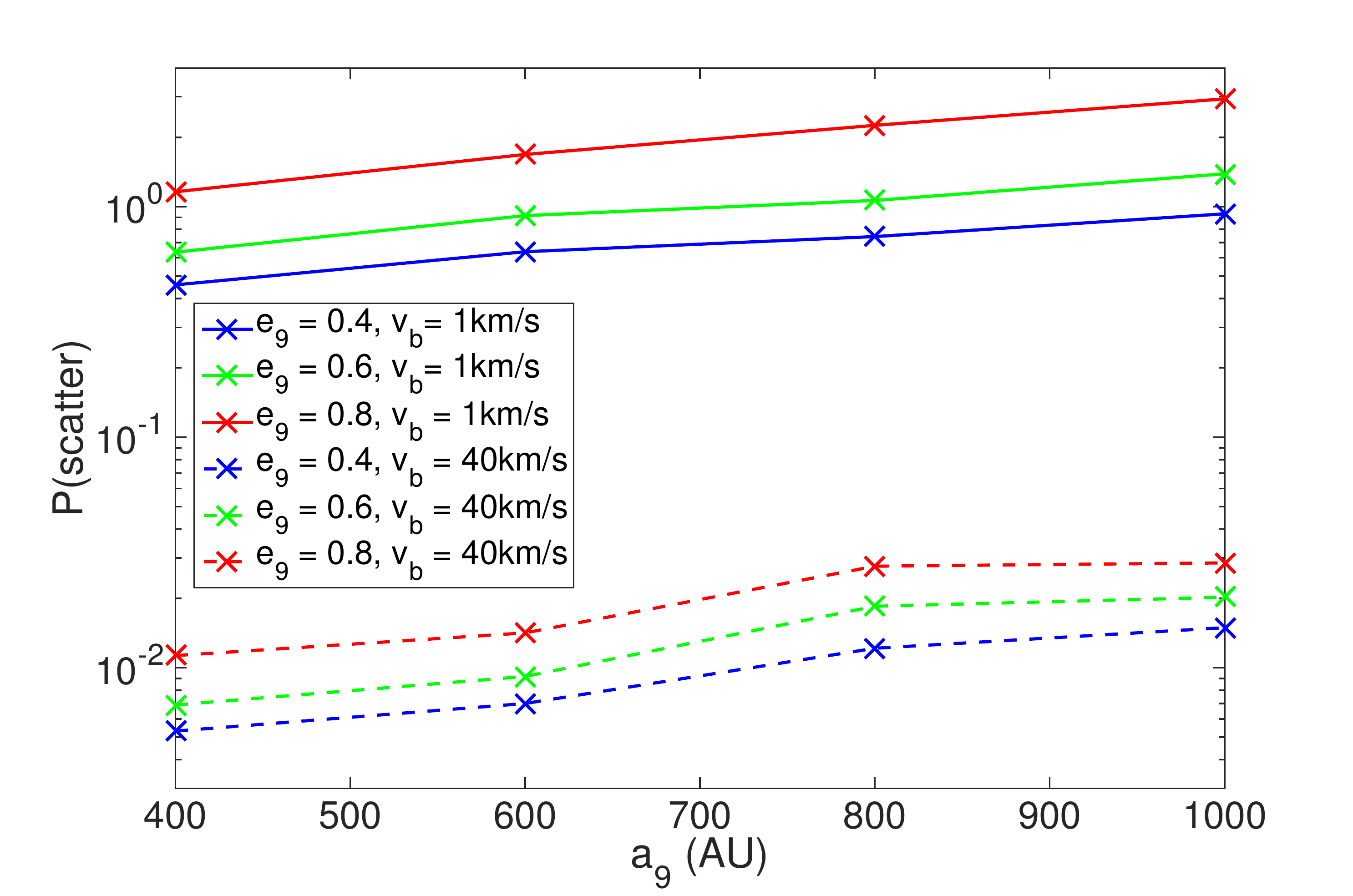} } } 
\figcaption{Scattering optical depth for removing Planet Nine from 
the Solar System, plotted versus semimajor axis. Families of curves
are shown for cluster environments ($v_b=1$ km/s; solid) and for the
field ($v_b=40$ km/s; dashed). For each velocity dispersion, results 
are shown for orbital eccentricities $e=0.4,0.6,$ and 0.8. } 
\label{fig:ejection} 
\end{figure}  

Next we consider the probability of removing Planet Nine from its
orbit. Here we consider a range of initial orbital elements, with
semimajor axis ${a_9}=400-1000$ AU and eccentricty $e_9=0.4-0.8$.
These ejection simulations are carried out for both the Solar birth
cluster ($v_b=1$ km/s) and the Solar neighborhood ($v_b=40$
km/s). Figure \ref{fig:ejection} shows the resulting optical depths
for ejecting Planet Nine from our Solar System, plotted versus initial
semimajor axis $a_9$. The scattering optical depths are of order unity
for the cluster environment (solid curves), but are lower,
$0.003-0.03$, for the field (dashed curves). Moreover, the interaction
rates increase with orbital eccentricity.  These results suggest that
lower eccentricities and late formation times are preferred.

Next we consider the capture of Planet Nine from another star, where
the planet has starting semimajor axis $a_0$.  Within the cluster
environment, the total capture optical depth and that for capture into                         
a viable Planet Nine orbit are $(P_T,P_9,a_0)=(0.0071,0.0002,50$AU),
$(0.013,0.0018,100$AU), $(0.017,0.0042,150$AU), and
$(0.021,0.0066,200$AU). The probability of successful Planet Nine
production through capture is thus $\lta1\%$. In addition to the low
probability of capture, another problem arises: The largest cross
sections for capture occur for systems where Planet Nine already has a
wide orbit. Even for $a_0=200$ AU, the probability of success is only
$\sim 0.6\%$, but the formation of Planet Nine in such a wide orbit is
problematic. It is thus unlikely that Planet Nine forms in a wide
orbit around another star and is subsequently captured, as this
scenario requires two low probability events to take place. 

Finally, for freely floating planets, the total capture optical depth
$P_T=0.019$, whereas the optical depth for capture into a viable
Planet Nine orbit $P_9=0.084$. These estimates assume that the density
and velocity of freely floating planets are comparable to
$\langle{n_\ast}\rangle$ and $v_b$. Since the capture cross sections
are roughly comparable for bound planets in wide orbits and
freely-floating planets, the resulting probabilities are also
comparable ($0.2\%$).

The cross sections in this paper are calculated for interactions with
binaries.  However, the correction factor for including single stars
is of order unity and has the form
\be{\cal{F}}=f_b+(1-f_b){\sigma_{ss}\over\sigma_{b}},\ee  
where $f_b\sim1/2$ is the binary fraction \citep{duchene}. The ratio
of cross sections for impinging single stars versus binaries
$\sigma_{ss}/\sigma_{b}\approx1/3$ \citep{gdawg}, so that
${\cal{F}}\approx0.67$.

\section{Conclusion} 
\label{sec:conclude} 

This paper determines cross sections for scattering interactions
between the proposed new Solar System member Planet Nine and passing
stars (see Table 1). Here we consider several different classes of
interaction events and estimate probabilities for their occurence.

First, the extreme orbit inferred for Planet Nine can be produced
(from an initially closer orbit) via scattering interactions with
passing binaries in the Solar birth cluster. The cross sections for
producing the required orbit increase with starting semimajor axis and
the corresponding probability of producing such an orbit is $\sim10\%$
(Figure \ref{fig:formation}). Such a high success rate requires the
initial orbit to be rather wide, $a_0=200-400$ AU.

Next we determine cross sections for ejecting Planet Nine from the
Solar System, where the planet starts in its current (inferred) orbit.
For both the Solar birth cluster and the field, cross sections are
calculated for a range of possible orbital elements $(a_9,e_9)$.  With
these cross sections, Planet Nine has a scattering optical depth of
order unity (Figure \ref{fig:ejection}) if the Sun resides within its
birth cluster for $(\Delta{t})\approx100$ Myr. More specifically, 
\be{P}=0.78\left({\sigmaint\over5\times10^6{\rm{AU}}^2}\right)\left({\langle{n_\ast}\rangle\over100{\rm pc}^{-3}}\right)
\left({v_b\over1{\rm km/s}}\right)\left(\Delta{t}\over100{\rm Myr}\right)\left({{\cal{F}}\over0.67}\right)\,,\ee
where we have used a representative value for the cross section; the 
factor $\cal{F}$ includes the correction for single stars.  The
corresponding scattering optical depth $P\approx0.002-0.02$ for
interactions in the field (Figure \ref{fig:ejection}). The longer
residence time in the field nearly (but not quite) compensates for 
the lower stellar density and smaller cross section. These results 
indicate that Planet Nine is most likely produced (either formed or
placed in its orbit) after the Sun leaves its birth cluster or near
the end of its residence time there. The capture of a free-floating
planet naturally occurs during cluster dispersal, but the
probabilities remain low.

Given that the current collection of Solar System planets is not
necessarily the original line-up, we calculate cross sections for
capturing Planet Nine.  In one scenario, the planet starts with a
circular orbit around another star and is captured into its current
orbit about the Sun. The cross sections for these capture events are
an order of magnitude smaller than the ejection cross sections. In
another scenario, Planet Nine starts as a freely floating planet and
is captured by our Solar System; these events take place with
slightly lower probability than capture from a bound orbit. Moreover,
capture events are most likely to take place while the Solar System is
leaving its birth aggregate, equivalently, when the cluster disperses
\citep{perets}. If the Sun captures Planet Nine through either
channel, the final-state orbital elements often fall in the range
necessary to align the Kuiper Belt objects (Figure
\ref{fig:aecapture}).  However, the probability of successful Planet
Nine capture is only $\lesssim1\%$.

The results of this paper constrain the dynamical history and possible
formation scenarios for Planet Nine. These scattering simulations
indicate that survival of Planet Nine is possible but not guaranteed.
The ejection cross sections increase (almost linearly) with semimajor
axis, so that smaller orbits favor survival and are therefore
preferred. This work also suggests that production of the required
orbit is somewhat problematic: Capture events can produce the
right orbital elements, but the overall cross sections for capture are
quite low.  Scattering Planet Nine into its current orbit from a
smaller initial orbit (within the Solar System) and capturing it from
a freely-floating state are also possible, but such events are less
likely than ejecting the planet. The formation scenario for Planet
Nine thus represents an important challenge for future investigation.

\section*{Acknowledgements}

We thank Konstantin Batygin, Juliette Becker, Hagai Perets, and Scott
Treamine for useful conversations, and the referee for useful
suggestions.


\begin{thebibliography}

\bibitem[Adams(1846)]{jcadams} 
Adams, J. C. 1846, MNRAS, 7, 149 

\bibitem[Adams(2010)]{adams2010} 
Adams, F. C. 2010, ARA\&A, 48, 47 

\bibitem[Adams \& Laughlin(2001)]{al2001}
Adams, F. C., \& Laughlin, G. 2001, Icarus, 150, 151 

\bibitem[Adams \& Laughlin(2003)]{al2003}
Adams, F. C., \& Laughlin, G. 2003, Icarus, 163, 290 

\bibitem[Adams et al.(2006)]{adams2006} 
Adams, F. C., Proszkow, E. M., Fatuzzo, M., \& Myers, P. C.
2006, ApJ, 641, 504 

\bibitem[Batygin \& Brown(2016)]{phattie} 
Batygin, K., \& Brown, M. E. 2016, AJ, 151, 22 

\bibitem[Binney \& Tremaine(2008)]{bintrem} 
Binney, J., \& Tremaine, S. 2008, Galactic Dynamics 
(Princeton: Princeton Univ. Press) 

\bibitem[Bovy et al.(2012)]{bovy} 
Bovy, J., Rix, H.-W., \& Hogg, D. W. 2012, ApJ, 753, 148  

\bibitem[Davis et al.(1984)]{nemesis} 
Davis, M., Hut, P., \& Muller, R. A. 1984, Nature, 308, 715 

\bibitem[de la Fuente Marcos \& de la Fuente Marcos(2014)]{dlfm}  
de la Fuente Marcos, C., \& de la Fuente Marcos, R. 2014, MNRAS, 443, L59  

\bibitem[Dodson-Robinson et al.(2009)]{dodson} 
Dodson-Robinson, S. E., Veras, D., Ford, E. B., \& Beichman, C. A.
2009, ApJ, 707, 79 

\bibitem[Duch{\^e}ne \& Kraus(2013)]{duchene} 
Duch{\^e}ne, G., \& Kraus, A. 2013, ARA\&A, 51, 269 

\bibitem[Duquennoy \& Mayor(1991)]{duque} 
Duquennoy, A., \& Mayor, M. 1991, A\&A, 248, 485

\bibitem[Galle(1846)]{galle} 
Galle, J. G. 1846, MNRAS, 7, 153 

\bibitem[J{\'i}lkov{\'a} et al.(2015)]{jilkova} 
J{\'i}lkov{\'a}, L., Portegies Zwart, S., Pijloo, T., \& Hammer, M.
2015, MNRAS, 453, 3157

\bibitem[Kenyon \& Bromley(2004)]{kenyon} 
Kenyon, S. J., \& Bromley, B. C. 2004, AJ, 128, 1916 

\bibitem[Lada \& Lada(2003)]{ladalada} 
Lada, C. J., \& Lada, E. A. 2003, ARA\&A, 41, 57

\bibitem[LeVerrier(1846)]{lever} 
Le Verrier, U. 1846, French Academy of Science 

\bibitem[Li \& Adams(2015)]{gdawg} 
Li, G., \& Adams, F. C. 2015, MNRAS, 448, 344 

\bibitem[Malmberg et al.(2007)]{malm2007} 
Malmberg, D., de Angeli, F., Davies, M. B., Church, R. P., Mackey, D.,
\& Wilkinson, M. I. 2007, MNRAS, 378, 1207

\bibitem[Malmberg et al.(2011)]{malm2011} 
Malmberg, D., Davies, M. B., \& Heggie, D. C. 2011, 
MNRAS, 411, 859 

\bibitem[McKee et al.(2015)]{mckee} 
McKee, C. F., Parravano, A., \& Hollenbach, D. J. 2015, 
ApJ, 814, 13  

\bibitem[Nordstrom et al.(2004)]{nordstrom} 
Nordstr{\"o}m, B., Mayor, M., Andersen, J., Holmberg, J., Pont, F.,
J{\o}rgensen, B. R., Olsen, E. H., Udry, S., \& Mowlavi, N. 2004, A\&A,
418, 989

\bibitem[Perets \& Kouwenhoven(2012)]{perets} 
Perets, H. B., \& Kouwenhoven, M.B.N. 2012, ApJ, 750, 83 

\bibitem[Pfalzner(2013)]{pfalzner}
Pfalzner, S. 2013, A\&A, 549, 82

\bibitem[Pfalzner et al.(2015)]{pfalznerrev}
Pfalzner, S., Davies, M. B., Gounelle, M., Johansen, A., M{\"u}nker, C.,
Lacerda, P., Portegies Zwart, S., Testi, L., Trieloff, M., \& Veras, D.
2015, Phys. Scripta, 90, 068001

\bibitem[Porras et al.(2003)]{porras}
Porras, A., et al. 2003, AJ, 126, 1916

\bibitem[Portegies Zwart(2009)]{zwart}
Portegies Zwart, S. F. 2009, ApJL, 696, L13

\bibitem[Proszkow \& Adams(2009)]{proszkow} 
Proszkow, E.-M., \& Adams, F. C. 2009, ApJS, 185, 486  

\bibitem[Tombaugh(1996)]{tombaugh} 
Tombaugh, C. W. 1996, Completing the Inventory of the Solar System,
ASP Conf. Series 107, 157 

\bibitem[Trujillo \& Sheppard(2014)]{ts2014} 
Trujillo, C. A., \& Sheppard, S. S. 2014, Nature, 507, 471 

\bibitem[Veras et al.(2009)]{veras} 
Veras, D., Crepp, J. R., \& Ford, E. B. 2009, ApJ, 696, 1600 

\end{thebibliography}
\end{document}